\documentstyle[epsfig,aps,prb,multicol]{revtex}
\tightenlines
\begin{document}
\def\beq{\begin{equation}}
\def\eeq{\end{equation}}
\def\beqa{\begin{eqnarray}}
\def\eeqa{\end{eqnarray}}
\newcommand{\com}[2]{{\mbox{\Large(}}^{#1}_{#2}{\mbox{\Large)}}}
\draft
\title{Monte Carlo simulation of nonlinear Couette flow in a dilute gas}
\author{Jos\'e Mar\'{\i}a Montanero\cite{chema}}
\address{Departamento de Electr\'onica e Ingenier\'{\i}a Electromec\'anica,\\
Universidad de Extremadura, E-06071 Badajoz, Spain}
\author{Andr\'es Santos\cite{andres} and Vicente Garz\'o\cite{vicente}}
\address{Departamento de F\'{\i}sica, Universidad de Extremadura,\\
E-06071 Badajoz, Spain}
\date{\today}
\maketitle

\begin{abstract}
The Direct Simulation Monte Carlo method is applied to solve the Boltzmann
equation in the steady planar Couette flow for Maxwell molecules and hard
spheres.
Nonequilibrium boundary conditions based on the solution of the
Bhatnagar-Gross-Krook (BGK) model for the Couette flow are employed to diminish the influence of
finite-size effects.
Non-Newtonian properties are characterized by five independent generalized transport
coefficients: a viscosity function, a thermal conductivity function, two
viscometric functions, and a cross coefficient measuring the heat flux
orthogonal to the thermal gradient. These coefficients depend
nonlinearly on the shear rate.
The simulation results are compared with theoretical predictions given by
the Grad method and the BGK and the ellipsoidal statistical (ES) models.
It is found that the kinetic models present a good agreement with the
simulation, especially in the case of the ES model, while the Grad method is
only qualitatively reliable for the momentum transport.
In addition, the velocity distribution function is also measured and
compared with the BGK and ES distributions.

\end{abstract}
\section{Introduction}
\label{sec1}

One of the most interesting states for analyzing transport processes far from
equilibrium is the steady planar Couette flow. The physical situation
corresponds to a system enclosed between two infinite, parallel plates in
relative motion and, in general, kept at different temperatures. These boundary
conditions lead to combined heat and momentum transport. If $x$ and $y$ denote
the coordinates parallel to the flow and orthogonal to the plates,
respectively,
then the corresponding steady hydrodynamic balance equations are
\beq
\label{1.1}
\frac{\partial P_{xy}}{\partial y}= \frac{\partial P_{yy}}{\partial y}=0,
\eeq
\beq
\label{1.2}
P_{xy}\frac{\partial u_x}{\partial y}+\frac{\partial q_y}{\partial y}=0,
\eeq
where ${\bf u}=u_x {\bf {\widehat x}}$ is the flow velocity,
${\sf P}$ is the pressure tensor, and ${\bf q}=q_x {\bf {\widehat x}}+q_y
{\bf {\widehat y}}$  is the heat flux.
The presence of $q_y$ in Eq.\ (\ref{1.2}) indicates that a thermal
gradient $\partial T/\partial y$ is
induced by the velocity gradient, even if both
plates are kept at
the same temperature. The balance equations (\ref{1.1}) and (\ref{1.2}) do
not
constitute a closed set unless the dependence of
the pressure tensor and the heat flux on the hydrodynamic fields is known.
If the gradients are small, the fluxes ${\sf P}$ and ${\bf q}$ are
described by the Navier-Stokes (NS) constitutive relations, which in this
problem yield
\beq
\label{1.3}
P_{xx}=P_{yy}=P_{zz},\quad P_{xy}=-\eta_{0}
\frac{\partial u_x}{\partial y},
\eeq
\beq
\label{1.4}
q_x=0,\quad q_y=-\kappa_{0}\frac{\partial T}{\partial y},
\eeq
where $\eta_0$ and $\kappa_0$ are the NS shear viscosity and
thermal
conductivity coefficients, respectively.
As a consequence of the absence of normal stress differences in the NS
description, the hydrostatic pressure $p=(P_{xx}+P_{yy}+P_{zz})/3$ is a
constant, on account of the balance equation (\ref{1.1}).

Even in the linear regime described by the NS equations, one still needs to
know
the spatial dependence of the transport coefficients to obtain the
exact solution of the hydrodynamic equations.
The problem becomes tractable in the case of
a low density gas, where the state of the
system is completely specified by the velocity distribution function $f({\bf r},
{\bf v};t)$, which obeys the Boltzmann equation. \cite{C90}
A relevant dimensionless quantity in a dilute gas is the Knudsen number $\text{Kn}=\lambda/\ell_h$, defined as the ratio of the mean free 
path $\lambda$ to the scale length of the hydrodynamic gradients $\ell_h$. 
In many laboratory conditions, $\text{Kn}\ll 1$ and so the Boltzmann equation can be solved by means of the Chapman-Enskog method as an 
expansion of the distribution function in powers of the Knudsen number.\cite{CC70}
The zeroth order approximation leads to the Euler hydrodynamic equations, while the first order approximation yields the NS equations 
with explicit expressions for the transport coefficients $\eta_0$ and $\kappa_0$. The results show that the ratio $\eta_0/\kappa_0$ is a 
constant. Consequently, 
 it then follows from the NS hydrodynamic equations (\ref{1.1})--(\ref{1.4})
that the flow velocity profile is quasi-linear,
\beq
\label{1.4.1}
\eta_0\frac{\partial u_x}{\partial y}=\text{const},
\eeq
and  the temperature is quasi-parabolic,
\beq
\left(\kappa_0\frac{\partial}{\partial
y}\right)^2 T=-\frac{\kappa_0}{\eta_0}
\left(\eta_0\frac{\partial u_x}{\partial y}\right)^2=\text{const}.
\label{1.4.3}
\eeq
Note that the profile of $u_x$ is not strictly linear, due to the space
dependence of $\eta_0$ through the temperature. Analogously, the
temperature profile is not strictly quadratic. In fact, the specific form of
both profiles depends on the interaction potential under consideration. On
the other hand, from Eqs.\ (\ref{1.4.1}) and (\ref{1.4.3})  it is easy to
derive a nice result, namely that if the temperature $T$ is seen as a
function of $u_x$ rather than as a function of the coordinate space $y$,
then one has
\beq
\label{1.4.2}
\frac{\partial^2T}{\partial u_x^2}=-\frac{\eta_0}{\kappa_0}.
\eeq
This is a sort of nonequilibrium ``equation of state,'' according to which the temperature is a quadratic function of the
flow velocity. Moreover, the ``curvature'' of the profile is practically
universal, given the weak influence of the interaction potential on the
Prandtl number  $\text{Pr}=5k_B\eta_0/2m\kappa_0\simeq \frac{2}{3}$, where
$k_B$ is the Boltzmann constant and $m$ is the mass of a particle.

Since the mean free path is inversely proportional to the density, the Knudsen number, at a given value of the scale length $\ell_h$, 
increases as the gas becomes more rarefied. In general,
 when the Knudsen number is not small, the NS relations are not expected to hold
and the transport must be described by nonlinear constitutive equations. In
the special case
of Maxwell molecules (particles interacting via an $r^{-4}$ potential),
it has been shown \cite{MN80,TS95}
 that the Boltzmann equation admits a consistent
solution in the {\em nonlinear\/} Couette flow characterized by  a constant
pressure $p$ and
profiles similar to
those obtained in the NS regime, Eqs.\ (\ref{1.4.1})--(\ref{1.4.2}), except that
$\eta_0$ and $\kappa_0$ are replaced
 by a generalized shear
viscosity coefficient $\eta(a)=\eta_0 F_{\eta}(a)$ and a generalized thermal
conductivity coefficient $\kappa_{yy}(a)=\kappa_0 F_{\kappa}(a)$, respectively.
Here, $a=(\eta_0/p)\partial u_x/\partial y$ is a constant (dimensionless)
shear rate and $F_{\eta}$ and
$F_{\kappa}$ are nonlinear functions of $a$. In addition,
$P_{xx}\neq P_{yy} \neq P_{zz}$ and $q_x\neq 0$.  
In this problem, the hydrodynamic scale  length can be identified as $\ell_h\sim \sqrt{k_BT/m}(\partial u_x/\partial y)^{-1}$, while the 
mean free path is $\lambda\sim \sqrt{k_BT/m}(\eta_0/p)$. Thus, the reduced shear rate $a$ represents the Knudsen number in this problem, 
i.e. $a\sim \text{Kn}$. 
Henceforth, we will use the reduced shear rate $a$ to refer to the Knudsen number $\text{Kn}$.
The solution considered in Refs.\ \onlinecite{MN80,TS95} describes heat and
momentum transport for arbitrary velocity and thermal gradients in the {\em bulk\/}
domain, where boundary effects are negligible.  On the other hand,
the full nonlinear dependence of
$F_{\eta}(a)$ and
$F_{\kappa}(a)$ is not explicitly known, since it involves the infinite
hierarchy of moment equations.
Their knowledge is limited to super-Burnett order  and the result
is \cite{TS95} $F_\eta(a)=1-3.111 a^2$ and $F_\kappa(a)=1-7.259 a^2$.

Consequently, if one wants to get the transport properties for arbitrary values
of the shear rate and the thermal gradient, one must resort to
approximate schemes or to computer simulations.
In the first alternative, explicit expressions for the nonlinear transport
coefficients in the Couette flow have been obtained from exact solutions of the
Bhatnagar-Gross-Krook (BGK) model \cite{BSD87,KDSB89} and related models
\cite{GLH94,GLH97,MG98} for general
interactions, as well as
from the Grad method. \cite{RC97} In the simulation side,
Risso
and Cordero \cite{RC97} have recently studied the shear-rate dependence of
$F_{\eta}$
and $F_{\kappa}$  by means of molecular dynamics simulations of a hard disk gas.
Comparison between the different analytical results with those obtained from the
simulation shows that the predictions given by kinetic models are in
better
agreement than those given by the Grad method,
especially in the case of
the thermal conductivity. \cite{MG98} Nevertheless, given the difficulties
associated with molecular dynamics simulations to achieve large shear rates
in a dilute gas, the above comparison is restricted to a range of shear rates
for which non-Newtonian effects are hardly significant. For
instance, the
shear viscosity has only decreased around $10\%$ with respect to its Navier-Stokes
value for the largest value of the shear rate considered
by Risso and Cordero. \cite{RC97}
In order to overcome such difficulties and extend the range of values of $a$,
one may use the so-called
Direct Simulation Monte Carlo (DSMC) method, \cite{Bird94} which is known to
qualify as an efficient and accurate method to numerically solve the
Boltzmann equation.

The aim of this paper is to solve the Boltzmann equation by means
of the DSMC method for a gas subjected to the planar Couette flow.
The motivation of this work is twofold. On the one hand, we want
to test the reliability of the far from equilibrium results obtained from kinetic
models and the Grad method by making a comparison with the
Boltzmann solution in the case of Maxwell molecules, for which the form of the hydrodynamic profiles in the bulk
region (far from the boundaries) is known. We will determine not only
the hydrodynamic profiles but also the nonlinear transport
coefficients and the velocity distribution function. On the other
hand, we want to investigate whether the above results for a
system of Maxwell molecules extend to other interaction
potentials. This extension holds when the Boltzmann equation is
replaced by kinetic models where, in terms of a conveniently
scaled space variable, all the results are independent of the
interaction law. Thus, we will also solve numerically the Boltzmann equation
by the DSMC method for a hard-sphere gas.

Since we are interested in describing transport properties in the bulk region,
i.e., free from finite-size effects, we need to use appropriate boundary
conditions in the simulations. In the conventional boundary conditions,
\cite{RC97,TTKB98} the gas is
assumed to be enclosed between two baths {\em at equilibrium} in relative
motion and, in general, at different temperatures.
Under these conditions, a particle leaving the
system is formally replaced by a particle coming from the bath, so the
incoming velocity is sampled from a local equilibrium distribution. As a
consequence, there exists a mismatch between the velocity distribution function
of the reemitted particles and that of those particles of the gas located
near the wall
and moving along the same direction. In this case, in order to inhibit  the
influence of
boundary effects one needs to take very large systems (normal distance
between the plates much larger than the mean free path), what is not
practical from a
computational point of view.
To overcome this difficulty, a possibility is to assume that both baths
are out
of equilibrium in a state close to that of the actual gas. Since such a
state is not known ``a priori'', in this paper
 we assume that the state of the baths is described by the BGK
solution of the planar Couette flow. \cite{KDSB89} Although the
boundary effects are still unavoidable, we expect that the above
mismatch between reemitted and gas particles will be much smaller.
As a matter of fact, the use of these conditions has been shown to
be  appropriate to analyze bulk transport properties in the
special case of planar Fourier flow (both walls at rest).
\cite{MASG94}

The plan of the paper is as follows. In Sec.\ \ref{sec2} we give a brief
description of the planar Couette flow and a summary of the main results
obtained from the Boltzmann equation and
kinetic models. The boundary conditions used in the
simulations and the DSMC method are described in Sec.\ \ref{sec3}. Section
\ref{sec4} presents the main results of the paper, where
special attention is paid to  the nonlinear transport coefficients. A
comparison with
the analytical results derived from the BGK and the
ellipsoidal statistical (ES) models and from the Grad method is also carried
out. The comparison shows in general a good agreement of the kinetic models
with computer simulations, even for large shear rates. In addition,
 the velocity distribution function obtained from the simulation in the bulk
domain is compared with the ones given by the kinetic models. It
is shown again that the agreement is qualitatively good. We close the paper
in Sec.\ \ref{sec5} with some concluding remarks.

\section{Description of the problem and summary of theoretical results}
\label{sec2}

Let us consider a dilute gas. In this case, a kinetic description is sufficient
to characterize the state of the system  by means of the velocity distribution
function $f({\bf r},{\bf v};t)$. This distribution function obeys the nonlinear
Boltzmann equation, which in the absence of external forces is given by
\cite{C90}
\beq
\label{2.1}
\frac{\partial f}{\partial t}+ {\bf v}\cdot \nabla f=J[f,f]  ,
\eeq
where
\beq
\label{2.2}
J[f,f]=\int d{\bf v}_1 \int d\widehat {{\bf k}}\, g I(g,\widehat{{\bf k}})
\left[
f({\bf v}') f({\bf v}_1')-f({\bf v}) f({\bf v}_1)\right]
\eeq
is the collision operator. In this equation, $I(g,\widehat{{\bf k}})$ is the
differential cross section, $g\equiv|{\bf v}-{\bf v}_1|$ being the relative
velocity, and $({\bf v}',{\bf v}'_1)$ are precollisional velocities
yielding postcollisional velocities $({\bf v},{\bf v}_1)$. From the
distribution function, one may define the hydrodynamic quantities
\beq
\label{2.3}
n=\int d{\bf v}\, f  ,
\eeq
\beq
\label{2.4}
{\bf u}=\frac{1}{n}\int d{\bf v}\, {\bf v}\, f  ,
\eeq
\beq
\label{2.5}
\frac{3}{2}nk_BT=\frac{m}{2}\int d{\bf v} \,{\bf V}^2\, f  ,
\eeq
and  the momentum and heat fluxes
\beq
\label{2.6}
{\sf P}=m \int d{\bf v}\, {\bf V}{\bf V}\, f  ,
\eeq
\beq
\label{2.7}
{\bf q}=\frac{m}{2} \int d{\bf v}\, {\bf V}^2{\bf V}\, f  .
\eeq
Here, $n$ is the number density, ${\bf u}$ is the flow velocity, $T$ is the
temperature, ${\sf P}$ is the pressure tensor, ${\bf q}$ is the heat
flux, and ${\bf V}={\bf v}-{\bf u}$ is the peculiar velocity. In
addition, the equation of state is that of an ideal gas, i.e., $p=nk_BT$.

Most of the known solutions to Eq.\ (\ref{2.1}) for spatially
inhomogeneous states \cite{SG95} correspond to the special case of
Maxwell molecules, namely, a repulsive potential of the form
$V(r)\sim r^{-4}$. For this potential, the collision rate $gI(g,
\widehat{{\bf k}})$ is independent of the relative velocity and
this allows  the infinite hierarchy of  velocity moments to be
recursively solved in some specific situations. Furthermore, the
NS transport coefficients $\eta_0$ and $\kappa_0$ can be exactly
obtained from the Chapman-Enskog method. \cite{CC70} They are
given by
\beq
\label{2.8}
\eta_0=\frac{p}{\nu} , \quad
\kappa_0=\frac{15}{4}\frac{k_B}{m}\eta_0  ,
\eeq
where
$\nu=\theta n$, $\theta$ being an eigenvalue of the linearized
Boltzmann collision operator. \cite{McL89}

In this paper we are interested in studying the planar Couette flow for a dilute
gas. We consider a gas enclosed between two parallel plates
in relative motion and maintained at different temperatures.
Under these
conditions, the system is driven out of equilibrium by the combined action
of the velocity and thermal gradients along the  direction normal to the
plates.
 After a transient period, the gas is
expected to reach a steady state and the corresponding Boltzmann
equation reads
\beq
\label{2.9} v_y \frac{\partial}{\partial
y}f=J[f,f] ,
\eeq
where we have chosen the axis $y$ as the one
normal to the plates. In general, this equation must be solved
subjected to specific boundary conditions. Nevertheless, in the
same spirit as in the Chapman-Enskog method, \cite{CC70} one may
look for ``normal'' solutions in which all the space dependence of
the distribution function occurs through a functional dependence
on the hydrodynamic fields. The normal solution describes the
state of the gas in the hydrodynamic regime, namely, for times
much longer than the mean free time and for distances from the
walls much larger than the mean free path. As mentioned in the
Introduction, it has been proved \cite{MN80,TS95} that, in the
case of Maxwell molecules, Eq.\ (\ref{2.9}) admits an exact
solution corresponding to the planar Couette flow problem. This
solution belongs to the {\em normal\/} class, so that no explicit
boundary conditions appear. In other words, all the space
dependence of $f$ is given through the local density, the local
velocity, the local temperature, and their  gradients. The
solution is characterized by hydrodynamic profiles that are a
simple generalization of those predicted by the NS approximation,
Eqs.\ (\ref{1.4.1})--(\ref{1.4.2}), namely
\beq
p=\text{const} ,
\label{2.11}
\eeq
\beq
\frac{1}{\nu(y)}\frac{\partial u_x}{\partial y}
\equiv a=\text{const}  ,
\label{2.12}
\eeq
\beq
\label{2.13}
\left[\frac{1}{\nu(y)}\frac{\partial}{\partial y}\right]^2
T=-\text{Pr}\frac{2m}{k_B}\gamma(a)  .
\eeq
The dimensionless
parameter $\gamma(a)$ is a nonlinear function of the reduced shear
rate $a$ that, by construction, behaves as $\gamma\approx a^2/5$
in the limit $a\to 0$. Again, the temperature can be seen as a
quadratic function of the flow velocity, but now the coefficient
$\eta_0/\kappa_0$ appearing in Eq. (\ref{1.4.2}) is replaced by a
shear-rate dependent coefficient
$(\eta_0/\kappa_0)5\gamma(a)/a^2$. Furthermore, in this solution
the pressure tensor is independent of the thermal gradient and the
heat flux vector is linear in the thermal gradient, but all these
fluxes are nonlinear functions of the shear rate. This nonlinear
dependence can be characterized through five generalized transport
coefficients. First, the shear stress $P_{xy}$ defines a
generalized shear viscosity $\eta(a)$ as
\beq
\label{2.14}
P_{xy}=-\eta(a)\frac{\partial u_x}{\partial y}\equiv -\eta_0
F_{\eta}(a) \frac{\partial u_x}{\partial y}  .
\eeq
Analogously,
the component of
 the heat flux parallel to the thermal gradient defines a generalized
thermal conductivity coefficient $\kappa_{yy}(a)$:
\beq
\label{2.15}
q_{y}=-\kappa_{yy}(a)\frac{\partial T}{\partial y}\equiv -\kappa_0
F_{\kappa}(a) \frac{\partial T}{\partial y} .
\eeq
The dimensionless functions $F_\eta(a)$ and $F_\kappa(a)$ are the
most relevant
quantities of the problem. They are related to the function $\gamma(a)$ by
 $\gamma(a)=a^2F_\eta(a)/5F_\kappa(a)$.
Normal stress differences are different from zero and are measured by two
viscometric  functions $\Psi_{1,2}(a)$ defined by
\beq
\label{2.16}
\frac{P_{yy}-P_{xx}}{p}=\Psi_1(a) a^2 ,
\eeq
\beq
\label{2.17}
\frac{P_{zz}-P_{yy}}{p}=\Psi_2(a) a^2.
\eeq
Another interesting quantity related to the pressure tensor is the friction
 coefficient $\mu(a)=-P_{xy}/P_{yy}$. To NS order, we simply have $\mu(a)=a$.
 In the non-Newtonian regime, we generalize this coefficient as
 $\mu(a)=a F_\mu(a)$, where the friction function $F_\mu(a)$ is
\beq
F_\mu(a)=\frac{F_\eta(a)}{1-[\Psi_2(a)-\Psi_1(a)]a^2/3}.
\label{2.17bis}
 \eeq
Finally, there exists a nonzero component of the heat
flux orthogonal to the thermal gradient given by
 \beq
\label{2.18}
q_{x}=-\kappa_{xy}(a)\frac{\partial}{\partial y}T\equiv -\kappa_0
\Phi(a) a \frac{\partial}{\partial y}T .
\eeq
The three functions $\Psi_{1,2}(a)$ and $\Phi(a)$  are generalizations of
Burnett coefficients.
In fact, $\Psi_1(0)=-14/5$, $\Psi_2(0)=4/5$, and $\Phi(0)=-7/2$ for Maxwell molecules
and for hard spheres in the first Sonine
approximation.\cite{CC70} The
 determination of the nonlinear transport coefficients $F_\eta$, $F_\kappa$,
$\Psi_{1,2}$, and $\Phi$ would imply the solution of an infinite hierarchy
that cannot be solved in a recursive way.
\cite{MN80} This hierarchy can only be solved step by step
when one performs a perturbation expansion in powers of the shear rate.
In fact, Tij and Santos \cite{TS95}
determined the solution up to  super-Burnett order:
\beq
F_\eta(a)=1-3.111 a^2+{\cal O}(a^4),
\label{2.19}
\eeq
\beq
F_\kappa(a)=1-7.259 a^2+{\cal O}(a^4).
\label{2.20}
\eeq
From Eqs.\ (\ref{2.19}) and (\ref{2.20}) it follows that
$\gamma(a)=(a^2/5)[1+4.148 a^2+{\cal O}(a^4)]$.
In addition, $F_\mu (a)=1-1.911 a^2+{\cal O}(a^3)$.

Although the above analyses are valuable, they have two
main limitations. On the one hand, they are restricted to the special case
of Maxwell molecules. For other interaction potentials (e.g., hard spheres),
the collisional moments involve all the moments of the distribution function
and, as a consequence, the hydrodynamic profiles (\ref{2.11})--(\ref{2.13})
are not {\em strictly\/} true.
On the other hand, even for Maxwell molecules, the above perturbative solution
is not useful for finite shear rates.
These two limitations can be overcome,
in the context of analytical methods, by introducing
additional approximations, such as the Grad method, \cite{RC97} or by
describing the system by means of kinetic models.
\cite{BSD87,GLH94,GLH97,MG98}
In these approaches, one looks for a solution having the same hydrodynamic profiles as in the case of the Boltzmann equation, cf.\ Eqs.\ 
(\ref{2.11})--(\ref{2.13}). As before, this solution describes the properties of the system in the bulk region, which is insensitive to 
the details of the boundary conditions.
From the Grad method and from the BGK and ES kinetic models, one explicitly gets the full nonlinear
shear-rate dependence of the transport coefficients in a non-perturbative
way. Moreover, the results are {\em universal\/}, namely the functions
$F_\eta(a)$, $F_\kappa(a)$, \ldots  are independent of the interaction
potential, provided that the reduced shear rate is defined as in Eq.\
(\ref{2.12}) with $\nu=p/\eta_0$.

The thirteen-moment Grad method consists of replacing the actual
distribution by \beq f\to
f_L\left\{1+\frac{m}{n(k_BT)^2}\left[\left(\frac{mV^2}{5k_BT}-1\right){\bf
V}\cdot {\bf
q}+\frac{1}{2}\left(P_{ij}-p\delta_{ij}\right)V_iV_j\right]\right\},
\label{2.23} \eeq where \beq f_L=n\left(\frac{m}{2\pi
k_BT}\right)^{3/2}\exp\left(-\frac{mV^2}{2k_BT}\right)
\label{2.24} \eeq is the local equilibrium distribution function.
When the approximation (\ref{2.23}) is inserted into the Boltzmann
equation (\ref{2.1}) and  velocity moments are taken, one gets a
closed set of equations for $n$, ${\bf u}$, ${\sf P}$, and ${\bf
q}$. According to the geometry of the planar Couette flow, there
are eight independent moments instead of the original thirteen
moments appearing in Eq.\ (\ref{2.23}). Risso and Cordero
\cite{RC97,RC98} found that the set of independent moment
equations, neglecting nonlinear terms in the fluxes, admits a
solution consistent with the profiles (\ref{2.11})--(\ref{2.13}).
In addition, they obtained explicit expressions for the transport
coefficients introduced above as nonlinear functions of the
reduced shear rate $a$. Those expressions are displayed in the
Appendix.

Now we consider the kinetic model approach. The basic idea is to
replace the detailed Boltzmann collision operator by a much
simpler term that otherwise retains the main physical properties
of the original operator $J[f,f]$. The most familiar choice is the
BGK model, \cite{C90}
\beq
J[f,f]\to -\nu(f-f_L),
\label{2.25}
\eeq
where $\nu$ is an effective collision frequency
that depends on the temperature, according to the interaction
potential. The NS transport coefficients of the BGK model are
$\eta_0=p/\nu$ and $\kappa_0=5pk_B/2m\nu$. Thus the BGK model has
the drawback that it predicts an incorrect value for the Prandtl
number, $\text{Pr}=1$. This is a consequence of the fact that
$\nu$ is the  only adjustable parameter in the model. This
deficiency is corrected by the so-called ellipsoidal statistical
(ES) model, \cite{C90} in which case
\beq
J[f,f]\to -\zeta
(f-f_R),
\label{2.26}
\eeq
where $\zeta$ is again an effective
collision frequency and the reference function $f_R$ is
\beq
f_R=n\pi^{-3/2}(\det {\sf A})^{-1/2}\exp\left(-{\sf A}^{-1}:{\bf
VV}\right),
\label{2.27}
\eeq
where
$A_{ij}=(2k_BT/m)\text{Pr}^{-1}\delta_{ij}-2(\text{Pr}^{-1}-1)P_{ij}/mn$.
The NS coefficients are now $\eta_0=p/(\zeta \text{Pr}^{-1})$ and
$\kappa_0=5pk_B/2m\zeta$. If, as in Eq.\ (\ref{2.8}), we define a
collision frequency $\nu=p/\eta_0$, then $\nu=\zeta
\text{Pr}^{-1}$ in the ES model. Note that the ES model reduces to
the BGK model if we formally make $\text{Pr}=1$. Therefore, the ES
model can be seen as an extension of the BGK model to account for
the correct Prandtl number, which can be particularly relevant in
the Couette flow where heat transport and momentum transport are coupled.
The results derived from the BGK and the ES models for the Couette
flow problem are also given in the Appendix. Apart from obtaining
the transport properties, the velocity distribution function can
be explicitly written. In particular, the BGK distribution is
given by \cite{KDSB89}
\beqa
f({\bf r},{\bf v})&=& n
\left(\frac{m}{2\pi
k_BT}\right)^{3/2}\frac{2\alpha(1+\alpha)^{3/2}}{\epsilon |\xi_y|}
\int_{t_0}^{t_1} dt\, \left[2t-(1-\alpha)t^2\right]^{-5/2}
\nonumber\\
& & \times \exp
\left\{-\frac{2\alpha}{1+\alpha}\frac{1-t}{\epsilon \xi_y}-
\frac{1+\alpha}{2t-(1-\alpha)t^2}\left[\left(\xi_x+\frac{2a\alpha}{1+\alpha}
\frac{1-t}{\epsilon}\right)^2+\xi_y^2+\xi_z^2\right]\right\}  .
\label{2.28}
\eeqa
Here, $(t_0,t_1)=(0,1)$ if $\xi_y>0$ and
$(t_0,t_1)=[1,2/(1-\alpha)]$ if $\xi_y<0$. Besides,
$\bbox{\xi}\equiv(m/2k_BT)^{1/2} {\bf V}$,
\beq
\label{2.29}
\alpha=\frac{\epsilon}{\left(\epsilon^2+8\gamma\right)^{1/2}}  ,
\eeq
and
\beq
\label{2.30}
\epsilon=\left(\frac{2k_BT}{m}\right)^{1/2}\frac{1}{\nu}
\frac{\partial}{\partial y} \ln T
\eeq
is a (local) reduced
thermal gradient. Equation (\ref{2.28}) shows that the
distribution function is a highly nonlinear function of the
reduced gradients $a$ and $\epsilon$.

In Ref.\ \onlinecite{MG98} a comparison between  the analytical results
obtained from the Grad method and the BGK and ES models with those obtained
from molecular
 dynamics
simulations \cite{RC97} for hard disks was carried out. It was found that
the three theories reproduced quite well the simulation data for $F_\eta$,
but the Grad method failed for $F_\kappa$ and $\Phi$. The latter quantity
was reproduced better by the ES model than by the BGK model.
Notwithstanding this, more definite conclusions could not be reached because
the simulation data were restricted to rather small shear rates, namely
$a\lesssim 0.2$. In this range of shear rates, non-Newtonian effects are not
especially significant. In addition, although the molecular dynamics
simulations correspond to a very dilute gas (area fraction $\phi\approx
1\%$), the collisional contributions to the transport coefficients (absent
in a Boltzmann description) are not {\em strictly\/} zero. Finally, as a
minor point, conclusions drawn in the context of two-dimensional systems
should not be extrapolated without caution to the more realistic case of
three dimensions.
As said in the Introduction, the aim of this paper is to solve numerically
the Boltzmann equation by means of the DSMC method for the planar Couette
flow and compare the results with the theoretical predictions. We will
consider three-dimensional systems of Maxwell molecules and hard spheres
subjected to shear rates as large as $a\simeq 1.2$.
In addition to the nonlinear transport coefficients, the comparison will be
extended to the level of the velocity distribution function itself.

\section{Boundary conditions and Monte Carlo simulation}
\label{sec3}
\subsection{Boundary conditions}
The goal now is to solve numerically the Boltzmann equation corresponding to
the planar Couette flow by using the successful DSMC method. \cite{Bird94}
The gas is enclosed between two parallel plates located at $y=0$ and $y=L$,
which are moving along the $x$-direction with velocities ${\bf
U}_0=U_0\widehat{\bf x}$ and ${\bf
U}_L=U_L\widehat{\bf x}$, respectively. In addition, they are kept
at temperatures $T_0$ and $T_L$, respectively.
In order to solve Eq.\ (\ref{2.9}), we need to impose the corresponding
boundary conditions.
They can be expressed in terms of the
kernels $K_{0,L}({\bf v}, {\bf v}')$ defined as follows. When a particle
with velocity ${\bf v}'$ hits the wall at $y=L$, the probability of being
reemitted with a velocity ${\bf v}$ within the range $d{\bf v}$ is
$K_{L}({\bf v}, {\bf v}')d{\bf v}$; the kernel $K_{0}({\bf v}, {\bf v}')$
represents the same but at $y=0$.
The boundary conditions are then \cite{DB77}
\beqa
\label{3.1}
\Theta(\pm v_y)|v_y| f(y=\{0,L\},{\bf v})&=&\Theta(\pm v_y)\int d{\bf v}'\,
|v_y'|K_{0,L}({\bf v}, {\bf v}')\nonumber\\
&&\times\Theta(\mp v_y')f(y=\{0,L\},{\bf v}',t).
\eeqa
In this paper we consider boundary conditions of complete accommodation with
the walls, so that $K_{0,L}({\bf v}, {\bf v}')=K_{0,L}({\bf v})$ does not
depend on the incoming velocity ${\bf v}'$ and can be written as
\beq
\label{3.2}
K_{0,L}({\bf v})=A_{0,L}^{-1}\Theta(\pm v_y)|v_y| \phi_{0,L}({\bf v}),\quad
A_{0,L}=\int d{\bf v}\,\Theta(\pm v_y)|v_y| \phi_{0,L}({\bf v}),
\eeq
where $\phi_{0,L}({\bf v})$ represents the probability distribution of a
fictitious gas in contact with the system at $y=\{0,L\}$. Equation
(\ref{3.2}) can then be interpreted as meaning that when a particle hits a
wall, it is absorbed and then replaced by a particle leaving the fictitious
bath. Of course, any choice of $\phi_{0,L}({\bf v})$ must be consistent with
the imposed wall velocities and temperatures, i.e.,
\beq
U_{0,L}=\int d{\bf v}\, v_x
\phi_{0,L}({\bf v}),
\label{3.3}
\eeq
\beq
k_BT_{0,L}=\frac{1}{3}m\int d{\bf v}\, ({\bf v}-{\bf
U}_{0,L})^2
\phi_{0,L}({\bf v}).
\label{3.4}
\eeq
The simplest and most common choice is that of a Maxwell-Boltzmann (MB)
distribution:
\beq
\phi_{0,L}^{\text{MB}}({\bf v})=\left(\frac{m}{2\pi k_BT_{0,L}}\right)^{3/2}
\exp\left[-\frac{m({\bf v}-{\bf U}_{0,L})^2}{2k_BT_{0,L}}\right].
\label{3.5}
\eeq
 Under these conditions, the system is
understood to be enclosed between two independent baths {\em at
equilibrium}.
While the conditions (\ref{3.5}) are adequate for analyzing boundary
effects, \cite{C90,W93} they are not very efficient when one is
interested  in obtaining the transport properties in the
bulk region.
In order to inhibit the influence of boundary effects, it is more convenient
to imagine that the two fictitious baths are in {\em nonequilibrium\/}
states resembling the state of the actual gas near the walls. More
specifically, we can assume that the fictitious gases are described by the
BGK equation, whose exact solution for the steady planar Couette flow
is given by Eq.\ (\ref{2.28}). In this case, the probability
distributions $\phi_{0,L}({\bf v})$ are
\beqa
\phi_{0,L}^{\text{BGK}}({\bf v})&=&
\pi^{-3/2}\frac{m}{
k_BT_{0,L}}\frac{\alpha_{0,L}(1+\alpha_{0,L})^{3/2}}{\epsilon_{0,L}
|v_y|} \int_{t_0}^{t_1} dt\, \left[2t-(1-\alpha_{0,L})t^2\right]^{-5/2}
\nonumber\\
& &
\times \exp
\left\{-\left(\frac{2k_BT_{0,L}}{m}\right)^{1/2}
\frac{2\alpha_{0,L}}{1+\alpha_{0,L}}\frac{1-t}{\epsilon_{0,L}
v_y}-
\frac{m}{2k_BT_{0,L}}\frac{1+\alpha_{0,L}}{2t-(1-\alpha_{0,L})t^2}
\right.\nonumber\\
&&\left.\times\left[\left(v_x-U_{0,L}+\frac{2a'\alpha_{0,L}}{1+\alpha_{0,L}}
\frac{1-t}{\epsilon_{0,L}}\right)^2+v_y^2+v_z^2\right]\right\}  ,
\label{3.6}
\eeqa
where $(t_0,t_1)=(0,1)$ if $v_y>0$ and $(t_0,t_1)=[1,2/(1-\alpha_{0,L})]$ if
$v_y<0$, and
$\alpha_{0,L}=\epsilon_{0,L}/[\epsilon_{0,L}^2+8\gamma_{\text{BGK}}(a')]^{1/2}$.
Here, $a'$ is the {\em estimated\/} value of the reduced shear rate,
as predicted by the BGK model for specific values of
the boundary parameters $U_{0,L}$ and $T_{0,L}$, and
$\gamma_{\text{BGK}}(a')$ is obtained from Eq.\ (\ref{A6}).
Given the values of the four independent boundary parameters $U_{0,L}$ and
$T_{0,L}$ (as well as the distance $L$), the shear rate $a'$ and the
local thermal gradients $\epsilon_{0,L}$ are fixed by the conditions
(\ref{2.11})--(\ref{2.13}). Therefore,
\beq
a'=\frac{U_L-U_0}{{\cal L}},
\label{3.7}
\eeq
\beq
\epsilon_{0,L}=\frac{1}{\cal L} \left(\frac{2k_BT_{0,L}}{m}\right)^{1/2}
\left[\frac{T_L-T_0}{T_{0,L}}\pm
\frac{m\gamma_{\text{BGK}}(a')\text{Pr}}{k_BT_{0,L}}{\cal L}^2\right].
\label{3.8}
\eeq
In these equations, ${\cal L}$ is related to the actual separation
$L$ between the plates through the nonlinear equation
\beq
L=\int_0^{\cal L}  \frac{ds}{\nu(s)},
\label{3.9}
\eeq
where $s$ is a variable in terms of which the temperature is a quadratic
function, namely
$T(s)=T_0[1+\epsilon_0(m/2k_BT_0)^{1/2}s-m\gamma_{\text{BGK}}(a')
\text{Pr}s^2/k_BT_0]$, and the $s$-dependence of the collision frequency
$\nu(s)$ appears only through the temperature (taking into account that
$p=\text{const}$).
The solution of the nonlinear set of equations (\ref{3.7})--(\ref{3.9})
gives $a'$, $\epsilon_{0,L}$, and ${\cal L}$ for any choice of $U_{0,L}$,
$T_{0,L}$, and $L$.
Nevertheless, from a practical point of view, it is more convenient to fix
$U_0$, $T_{0,L}$, $a'$, and $\epsilon_0$ as independent parameters. Without
loss of generality we take $U_0=0$ and $T_L=1$. In addition, we will choose
$\epsilon_0=0$. This implies that, if boundary effects were absent, the
temperature would have a maximum at the lower plate $y=0$. Thus,
\beq
U_L= a'{\cal L}, \quad
{\cal L}=\left[\frac{\Delta}{2\text{Pr}\gamma_{\text{BGK}}(a')}\right]^{1/2},
\label{3.10}
\eeq
\beq
\epsilon_L=-2\frac{\Delta}{\cal L}.
\label{3.11}
\eeq
In the above equations  $\Delta\equiv T_0-1$ and we have taken $m=1$ and
$k_B= \frac{1}{2}$.
Finally, the actual distance $L$ is given by Eq.\ (\ref{3.9}). However, from
the simulation point of view, it is more useful to express $L$ in terms of
the collision frequency $\overline{\nu}$ corresponding to the temperature
$T_L$ and the average density $\overline{n}$. In other words, instead of
Eq.\ (\ref{3.9}) we use
\beq
\overline{n}=\frac{1}{L}\int_0^L dy\, n(y)=\frac{1}{L}\int_0^{\cal L} ds
\frac{n(s)}{\nu(s)}
\label{3.12}
\eeq
to determine $L$. For the sake of concreteness, let us consider repulsive
potentials, for which $\nu=\overline{\nu}(n/\overline{n})(T/T_L)^\omega$, where
$\omega$ ranges from 0 (Maxwell molecules) to $\frac{1}{2}$ (hard spheres). In
that case,
\beq
L=\frac{1}{\overline{\nu}}\int_0^{\cal L}ds\, [T(s)]^{-\omega}.
\label{3.13}
\eeq
For Maxwell molecules, this simply reduces to $L={\cal L}/{\overline{\nu}}$,
while for hard spheres one has $L=({\cal
L}/{\overline{\nu}})\tan^{-1}(\sqrt{\Delta})/\sqrt{\Delta}$.
In summary, given the values of $a'$ and $\Delta$, the separation $L$ between
the plates and the velocity of the upper plate $U_L$ are uniquely
determined. In addition, the thermal gradient at the upper plate
$\epsilon_L$ is also determined, while the value at the lower plate is fixed
as $\epsilon_0=0$. The knowledge of these boundary parameters allows one to
obtain the distributions $\phi_{0,L}^{\text{BGK}}({\bf v})$, according to
Eq.\ (\ref{3.6}).

\subsection{The DSMC method}
Now, we briefly describe the numerical algorithm we have employed to solve
the Boltzmann equation by means of the so-called Direct Simulation Monte
Carlo (DSMC) method. \cite{Bird94}
In this method, the velocity
distribution function is  represented by the velocities $\{{\bf v}_i\}$
and positions $\{y_i\}$ of a sufficiently large number of particles $N$.
Given the geometry of the problem, the physical system is split into layers
of width $\delta y$, sufficiently smaller than the mean free path. The
velocities and coordinates are updated from time $t$ to time $t+\delta t$,
where the time step $\delta t$ is much smaller
than the mean free time, by applying a   streaming step followed by a
 collision step.
In the {\em streaming step}, the particles are moved
ballistically,
$y_i\to y_i+v_{iy}\delta t$. In addition, those
particles crossing the boundaries are reentered with velocities sampled from
the corresponding probability distribution $K_{0,L}({\bf v})$.
 Suppose that a particle $i$ crosses the lower plate
between times $t$ and $t+\delta t$, i.e., $y_i+v_{iy}\delta t<0$. Then,
regardless of the incoming velocity ${\bf v}_i$,  a new velocity
$\widetilde{\bf v}_i$ is assigned according to the
following rules.
First, a
velocity $\widetilde{\bf v}_i$ (with $\widetilde{v}_{iy}>0$) is sampled
with a probability proportional to
$|v_y|\phi_0^{\text{MB}}({\bf v})$.
If one is considering the MB boundary conditions, this
velocity is accepted directly. Otherwise,
the above acts as a ``filter'' to optimize
the acceptance-rejection procedure and the velocity $\widetilde{\bf v}_i$
is accepted with a probability proportional to the ratio
$\phi_0^{\text{BGK}}({\bf v})/\phi_0^{\text{MB}}({\bf v})$. If the velocity
is rejected, a new velocity $\widetilde{\bf v}_i$ is sampled and the process
is repeated until acceptance. The new position is assigned as
$\widetilde{v}_{iy}(\delta t+ y_i/{v}_{iy})$. The process is analogous in
the case of the upper plate.

The {\em
collision step\/} proceeds as follows
For each layer $\alpha$,
a pair of potential collision partners $i$ and $j$ are chosen at random with
equiprobability.
The collision between those particles  is then
accepted with a probability equal to the corresponding collision
rate times $\delta t$. For hard spheres, the collision rate is proportional
to the relative velocity $|{\bf v}_i-{\bf v}_j|$, while it is independent of
the relative velocity for Maxwell molecules (an angle cut-off is needed in
the latter case).
 If the collision is accepted, the scattering direction is randomly chosen
 according to the interaction law
 and
post-collisional velocities  are assigned to both particles, according to
the conservation of momentum and energy.
After the collision is processed or if the pair is rejected, the routine
moves again to the choice of a new pair until the required number of
candidate pairs  has been taken.

In the course of the simulations, the following ``coarse-grained'' local
quantities are computed. The number density in layer $\alpha$ is
\beq
n_\alpha=\overline{n}\frac{N_\alpha}{(N/L)\delta y}=\frac{\overline{n}
L}{N\delta y}\sum_{i=1}^N
\Theta_\alpha(y_i),
\label{3.14}
\eeq
where $\Theta_\alpha(y)$ is the characteristic function of layer $\alpha$,
i.e., $\Theta_\alpha(y)=1$ if $y$ belongs to layer $\alpha$ and is zero
otherwise.
Similarly, the flow velocity, the temperature, the pressure tensor, and the
heat flux of layer $\alpha$ are
\beq
{\bf u}_\alpha=\frac{1}
{N_\alpha}\sum_{i=1}^N
\Theta_\alpha(y_i) {\bf v}_i,
\label{3.15}
\eeq
\beq
k_BT_\alpha=\frac{p_\alpha}{n_\alpha}=\frac{m}{3N_\alpha}\sum_{i=1}^N
\Theta_\alpha(y_i) ({\bf v}_i-{\bf u}_\alpha)^2,
\label{3.16}
\eeq
\beq
{\sf P}_\alpha=\frac{L}{N\delta y}m\sum_{i=1}^N
\Theta_\alpha(y_i) ({\bf v}_i-{\bf u}_\alpha)({\bf v}_i-{\bf u}_\alpha),
\label{3.17}
\eeq
\beq
{\bf q}_\alpha=\frac{L}{N\delta y}\frac{m}{2}\sum_{i=1}^N
\Theta_\alpha(y_i) ({\bf v}_i-{\bf u}_\alpha)^2({\bf v}_i-{\bf u}_\alpha).
\label{3.18}
\eeq
From these quantities one can get local values of the gradients and of the
transport coefficients. For instance, the reduced shear rate is
\beq
a_\alpha=\frac{\overline{n}}{\overline{\nu}n_\alpha}\left(\frac{T_L}{T_\alpha}\right)^\omega
\frac{u_{\alpha+1,x}-u_{\alpha,x}}{\delta y}
\label{3.18.1}
\eeq
and the viscosity function is
\beq
F_{\eta,\alpha}=-\frac{P_{\alpha,xy}}{a_\alpha p_\alpha}.
\label{3.18.2}
\eeq

As said before, in the simulations we take units such that $m=1$,
$T_L=1$, and $k_B=\frac{1}{2}$. It remains to fix the time unit or, equivalently, the
length unit. The standard definition of mean free path in the case
of hard spheres is \cite{CC70}
\beq
\lambda=\frac{1}{\sqrt{2}
n\pi\sigma^2},
\label{3.19}
\eeq
where $\sigma$ is the diameter of
the spheres. The Navier-Stokes shear viscosity is (in the first
Sonine approximation)
$\eta_{0}=5(mk_BT/\pi)^{1/2}/16\sigma^2$. Consequently,
the effective collision frequency $\nu=p/\eta_{0}$ and the
mean free path $\lambda$ are related as
$\nu=(8/5\sqrt{\pi})(2k_BT/m)^{1/2}/\lambda$. As usual, we choose
the mean free path corresponding to the average density
$\overline{n}$ as the length unit. This in turn implies that
$\overline{\nu}=8/5\sqrt{\pi}\simeq 0.903$. For convenience, we
take the latter value for Maxwell molecules as well. The typical
values of the simulation parameters are $N= 2\times 10^5$ particles, a
layer width $\delta y=0.02$, and a time step $\delta t=0.003$.

The procedure  to measure the relevant quantities of the problem is
as follows.
First the values of the imposed shear rate $a'$ and the temperature
difference $\Delta$ are chosen. This choice fixes the system size $L$, as
well as the upper velocity $U_L$ and the upper thermal gradient
$\epsilon_L$, according to Eqs.\ (\ref{3.10}), (\ref{3.11}), and
(\ref{3.13}).
Starting from an equilibrium initial state with $T=T_0$, the system evolves
driven by the boundary conditions described before.
After a transient period (typically up to $t=25$), the system reaches a
steady state in which the quantities fluctuate around constant values.
In this state, the balance equations predict that the quantities $u_y$,
$P_{xy}$, $P_{yy}$, and $u_xP_{xy}+q_y$ are spatially uniform and this is
used in the simulations as a test to make sure that the steady state has
been achieved.
Once the steady state is reached, the local quantities
(\ref{3.14})--(\ref{3.18.2}) are averaged over typically 100 snapshots
equally spaced between $t=25$ and $t=55$, which corresponds to about 60 collisions per particle in the case of hard spheres. 

\subsection{Test of the numerical algorithm}
\begin{figure}
\begin{center}
\parbox{0.5\textwidth}{
\epsfxsize=\hsize \epsfbox{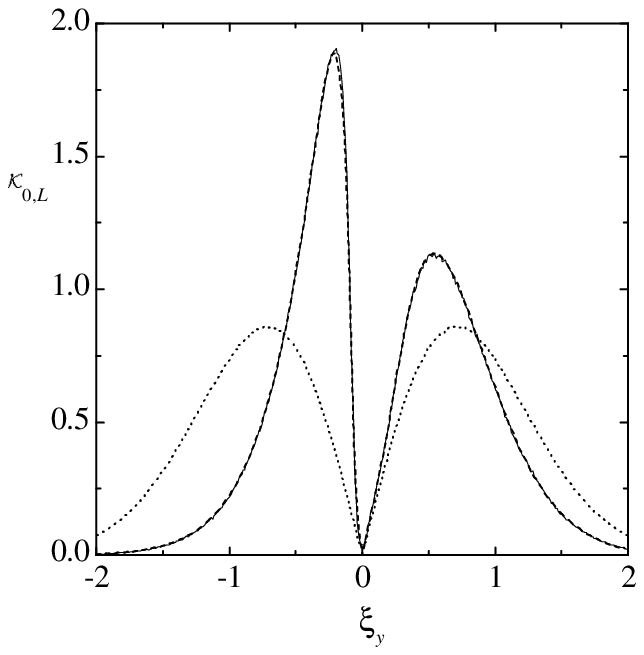}}
\caption{Plot of the marginal distribution function of particles  reemitted by the
walls. The solid line corresponds to the distribution measured in the
simulation by applying the BGK boundary conditions (with $a'=1$,
$\epsilon_0=0$, and $\epsilon_L=-3.15$), the dashed line is the theoretical
BGK distribution (which is hardly distinguishable from the solid line), and
the dotted line corresponds to the MB boundary conditions.
\label{fig1}
}
\end{center}
\end{figure}

\begin{figure}
\begin{center}
\parbox{0.5\textwidth}{
\epsfxsize=\hsize \epsfbox{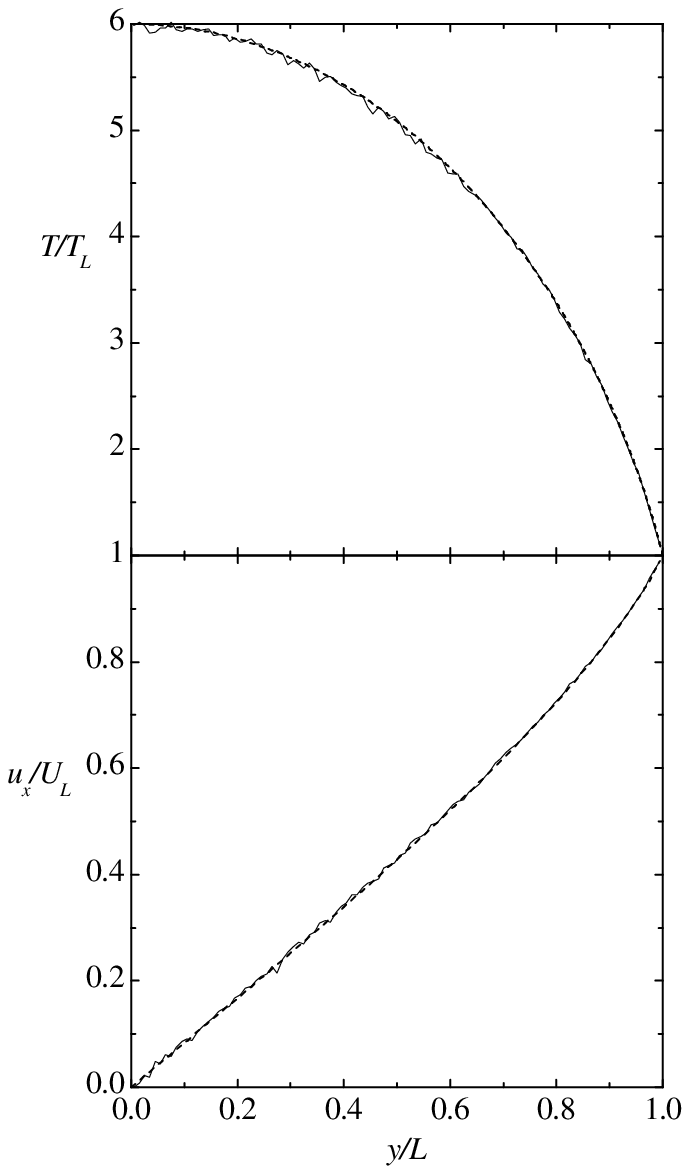}}
\caption{
Temperature and velocity profiles for hard spheres in the case $L=1.81$,
$U_L=3.17$, and $T_0=6$. The solid lines are results obtained from a DSMC
simulation of the BGK equation, while the dashed lines are the theoretical
BGK results.
 \label{fig2}}
\end{center}
\end{figure}

\begin{figure}
\begin{center}
\parbox{0.5\textwidth}{
\epsfxsize=\hsize \epsfbox{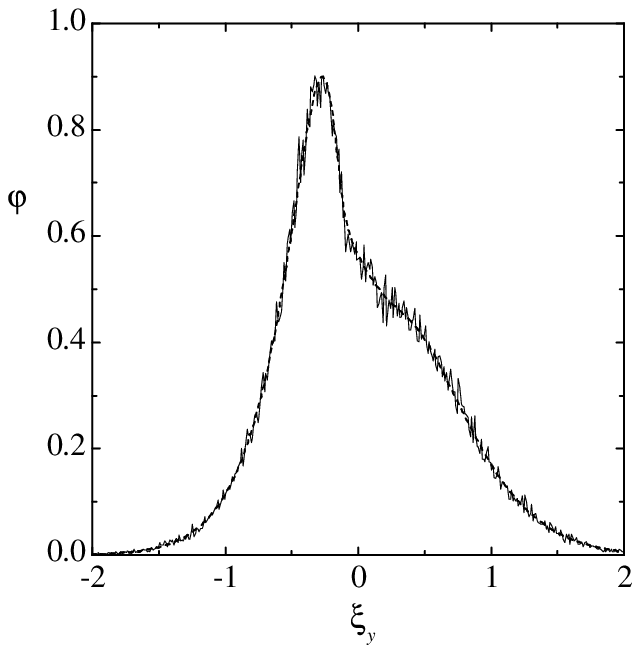}}
\caption{
Marginal velocity distribution function for hard spheres at $y=0.5 L$ in the
case $L=1.81$, $U_L=3.17$, and $T_0=6$. The solid line is
obtained from a DSMC simulation of the BGK equation, while the dashed line (which is hardly distinguishable from the solid line)
is the theoretical BGK distribution.
 \label{fig3}}
\end{center}
\end{figure}

Before closing this Section, it is worthwhile carrying out a test of the
reliability of the numerical method. To that end, we have simulated the BGK
equation by a DSMC-like method similar to the one described in Ref.\
\onlinecite{instab}.
If the boundary conditions are implemented correctly and the simulation
parameters  are well chosen, then the simulation results should agree with
the theoretical BGK predictions. We have checked that this indeed the case.
As an example, consider the hard-sphere situation  with $a'=1$ and $\Delta=5$.
This corresponds to  $\gamma_{\text{BGK}}=0.248$, $L=1.81$,
$U_L=3.17$, and $\epsilon_L=-3.15$, where in this case $\text{Pr}=1$.
Figure \ref{fig1} shows the marginal velocity distributions of particles
reemitted by the walls,
\beq
{\cal K}_{0,L}(\xi_y)=
\left(\frac{2k_BT_{0,L}}{m}\right)^{1/2}\int_{-\infty}^\infty dv_x
\int_{-\infty}^\infty dv_z\, K_{0,L}({\bf v}),
\label{3.19.3}
\eeq
as functions of $\xi_y=(m/2k_BT_{0,L})^{1/2} v_y$.
The case $\xi_y>0$ ($\xi_y<0$) corresponds to particles that are reemitted
from the lower (upper) plate. The agreement with the imposed distribution is
excellent. Note the strong asymmetry between the distributions corresponding
to both plates, in contrast to the symmetry of the MB
distributions obtained from (\ref{3.5}).
The  temperature and velocity profiles  are shown in Fig.\ \ref{fig2}.
The simulation values overlap, within statistical fluctuations, with the theoretical predictions.
Apart from the profiles, we have verified that the generalized transport
coefficients obtained from the fluxes also agree with the theory.
A more stringent test is provided in Fig.\ \ref{fig3}, where the marginal
distribution function
\beq
\varphi(\xi_y)=
\frac{1}{n}\left(\frac{2k_BT}{m}\right)^{1/2}\int_{-\infty}^\infty dv_x
\int_{-\infty}^\infty dv_z\, f({\bf v}), \quad
\xi_y=\left(\frac{m}{2k_BT}\right)^{1/2} v_y
\label{3.19.2}
\eeq
is plotted at the point $y/L=0.5$, which corresponds to a local
thermal gradient $\epsilon= -0.60$.
It is apparent again that an excellent agreement exists between
simulation and theory.

\section{Results}
\label{sec4}
This Section is devoted to a comparison between the simulation results for the Boltzmann equation
obtained by the simulation method described in the previous Section and the
theoretical predictions provided by the Grad method and the BGK and ES
kinetic models.
The comparison will be carried out at the levels of the transport
coefficients and the velocity distribution, both for Maxwell molecules and
hard spheres.
Before that, the hydrodynamic profiles obtained from the simulations by
using the two types of boundary conditions considered in Sec.\ \ref{sec3}
are presented.

\subsection{Hydrodynamic profiles}
\begin{figure}
\begin{center}
\parbox{0.5\textwidth}{
\epsfxsize=\hsize \epsfbox{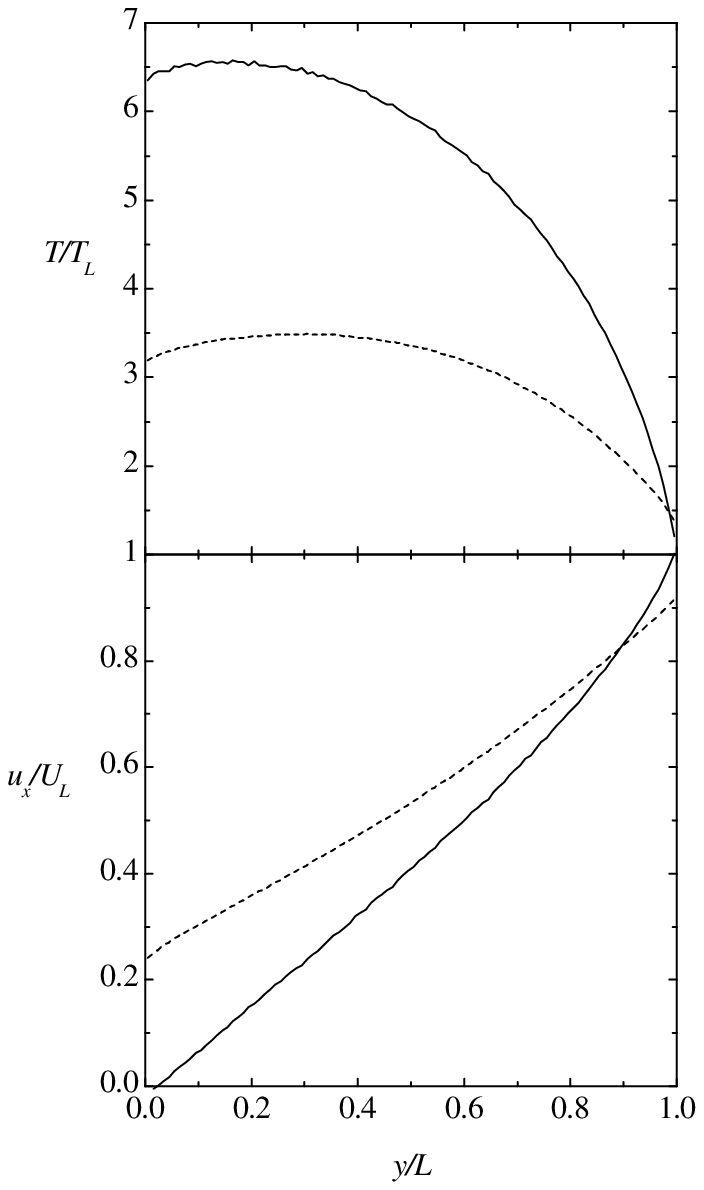}}
\caption{
Temperature and velocity profiles for Maxwell molecules, as obtained from
 DSMC simulations, in the case $L=4.68$, $U_L=3.90$, and $T_0=6$. The solid
 lines refer to the results obtained from the BGK boundary conditions and the dashed lines
refer to those obtained from the MB boundary conditions.
 \label{fig4}}
\end{center}
\end{figure}

\begin{figure}
\begin{center}
\parbox{0.5\textwidth}{
\epsfxsize=\hsize \epsfbox{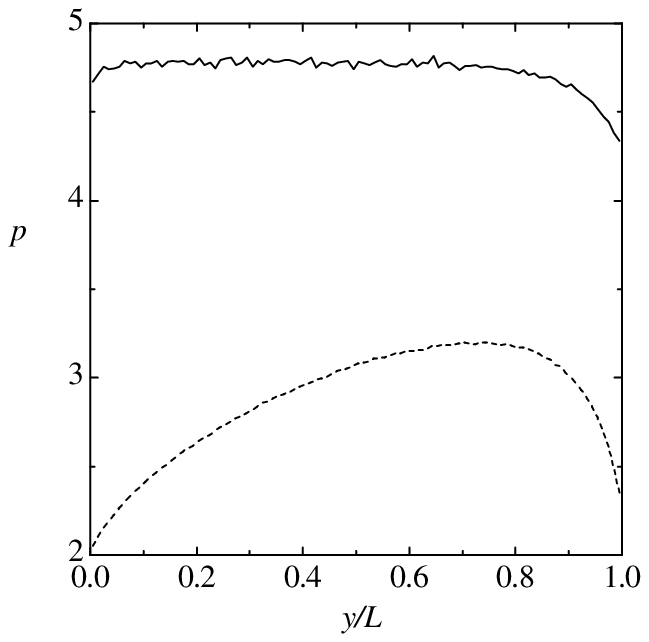}}
\caption{
Same as in Fig.\ \protect\ref{fig4}, but for the pressure (which is
measured in units of $\overline{n}k_BT_L$).
 \label{fig5}}
\end{center}
\end{figure}

\begin{figure}
\begin{center}
\parbox{0.5\textwidth}{
\epsfxsize=\hsize \epsfbox{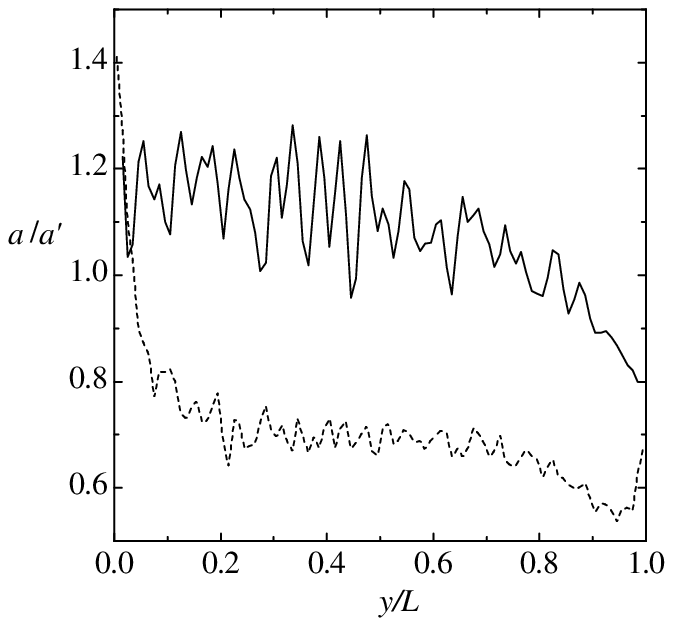}}
\caption{
Same as in Fig.\ \protect\ref{fig4}, but for the ratio $a/a'$ between the
actual shear rate, $a$, and the one imposed by the BGK boundary conditions,
$a'$.
 \label{fig6}}
\end{center}
\end{figure}

As said in Secs.\ \ref{sec1} and \ref{sec2}, the Boltzmann equation for
Maxwell molecules admits an exact solution characterized by Eqs.\
(\ref{2.11})--(\ref{2.13}).
This solution applies to the bulk region, i.e.,  the  region where boundary
effects are negligible. Obviously, in a simulation with a finite size of the
system, it is not possible to avoid boundary effects completely.
On the other hand, one can expect that the ``nonequilibrium'' boundary
conditions based on the BGK distribution, Eq.\ (\ref{3.6}), inhibit the
boundary effects, as compared with the conventional ``equilibrium'' boundary
conditions (\ref{3.5}).
We have confirmed that this is indeed the case. As an illustrative example,
let us consider  $\Delta=5$ and $a'=0.92$ for Maxwell molecules.
This corresponds to $\gamma_{\text{BGK}}=0.21$, $L=4.68$, $U_L=3.90$, and
$\epsilon_L=-2.37$.
Figure \ref{fig4} shows the temperature and velocity profiles as obtained by
using the MB and BGK boundary conditions.
It is apparent that the velocity slips and the temperature jumps at the
walls are much larger in the former case than in the latter.
Note that the maximum temperature is not exactly
located in the layer adjacent to the lower plate but it is slightly shifted.
It is interesting to remark that when plotting the temperature $T$ as a function of the flow velocity $u_x$, a parabolic curve is 
observed with both boundary conditions, as expected from Eqs.\ (\ref{2.12}) and (\ref{2.13}).
A more evident proof of the advantage of the BGK conditions is shown in
Fig.\ \ref{fig5}. Since the pressure is a constant in the exact solution
valid in the bulk, any deviation from $p=\text{const}$ can be attributed to
boundary effects. The pressure obtained with the BGK boundary conditions is
practically constant, except near the upper plate, while the one obtained
with the MB conditions is only nearly constant in a small region around $y/L\simeq 0.75$.
Finally, Fig.\ \ref{fig6} shows the ratio $a/a'$ between the actual (local)
shear rate $a$ measured in the simulations, cf. Eq.\ (\ref{3.18.1}), and the
imposed shear rate $a'$. The ratio is closer to 1 in the case of the BGK
boundary conditions than in that of the MB conditions. In addition, in the
former case the region where $a$ is practically constant extends to layers
closer to the lower plate.

In summary, the above example illustrates  that the BGK boundary conditions
are much more efficient than the MB ones to measure transport properties in
the bulk. Therefore, in what follows we will only consider the BGK
conditions.
In each case, we identify a bulk domain comprised between the layers $y=y_0$ and $y=y_1$ where $a\simeq
\text{const}$, $p\simeq \text{const}$, and $\gamma\simeq \text{const}$,
 and take averages of $a$,
$F_\eta$, $F_\kappa$, $\Psi_{1,2}$, and $\Phi$ over those layers. Typical values are $y_0/L\simeq 0.2$ and $y_1/L\simeq 0.8$.

\subsection{Nonlinear transport coefficients}
\begin{figure}
\begin{center}
\parbox{0.5\textwidth}{
\epsfxsize=\hsize \epsfbox{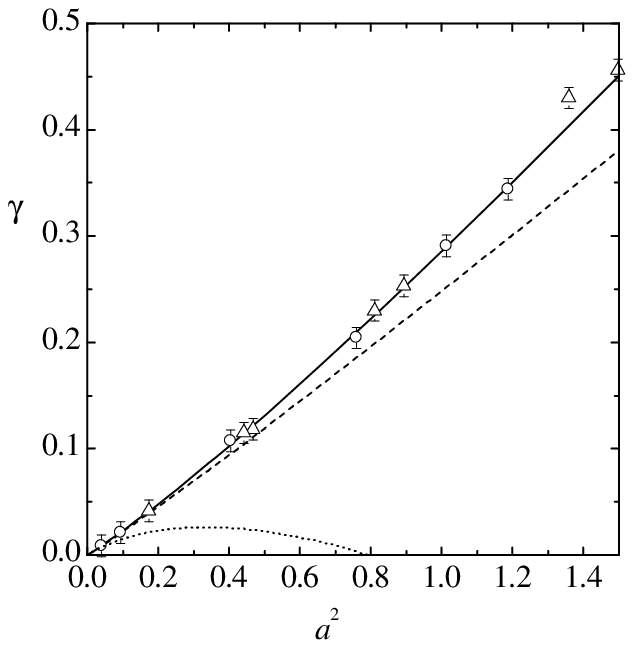}}
\caption{
Plot of the thermal curvature parameter $\gamma$ as a function of the shear
rate. The symbols are simulation data for Maxwell molecules
(circles) and for hard spheres (triangles), while the lines are the
theoretical predictions given by the ES kinetic model (solid line), the BGK
kinetic model  (dashed line), and the Grad method (dotted line).
 \label{fig7}}
\end{center}
\end{figure}

In this subsection we compare the simulation results for Maxwell molecules
and hard spheres obtained from the Monte Carlo simulations with the
(universal) predictions of the Grad method and the BGK and ES kinetic models.
As said in the Introduction, we are interested in situations where the Knudsen number is not small, so that nonlinear effects are 
important.\cite{MS96}
An interesting quantity in the {\em nonlinear\/} Couette flow is the
parameter $\gamma$, which is related to the curvature of the temperature
profile. Its shear-rate dependence is shown in Fig.\ \ref{fig7}.
A remarkable feature is that the simulation data for both interactions seem
to lie on a common curve. This indicates that, as predicted by the models,
the transport properties are hardly sensitive to the interaction potential,
provided that the quantities are conveniently scaled.
While the kinetic models exhibit a good quantitative (in the case of the ES
model) or qualitative (BGK model) agreement, the Grad method fails, except
for small shear rates.
Since $\gamma(a)=a^2 F_\eta(a)/5F_\kappa(a)$, one can interpret
$5\gamma(a)/a^2$ as an effective, shear-rate dependent Prandtl number
(relative to the usual $\text{Pr}$). This quantity is bounded between 1
for small shear rates and $5/3$ (in the BGK model) or $5/2$ (in the ES
model) in the limit of large shear rates. This is consistent with the fact
that the BGK model underestimates the value of $\gamma$.

\begin{figure}
\begin{center}
\parbox{0.5\textwidth}{
\epsfxsize=\hsize \epsfbox{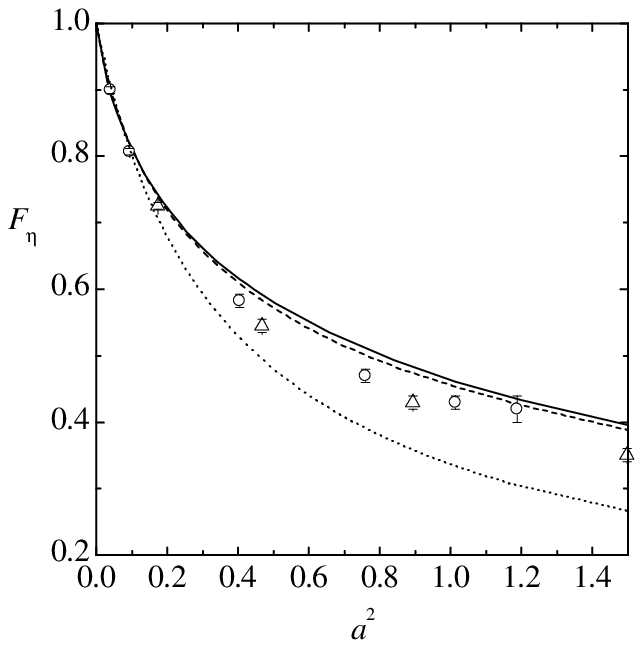}}
\caption{
Same as in Fig.\ \protect\ref{fig7}, but for the viscosity function
$F_\eta$.
\label{fig8}}
\end{center}
\end{figure}

\begin{figure}
\begin{center}
\parbox{0.5\textwidth}{
\epsfxsize=\hsize \epsfbox{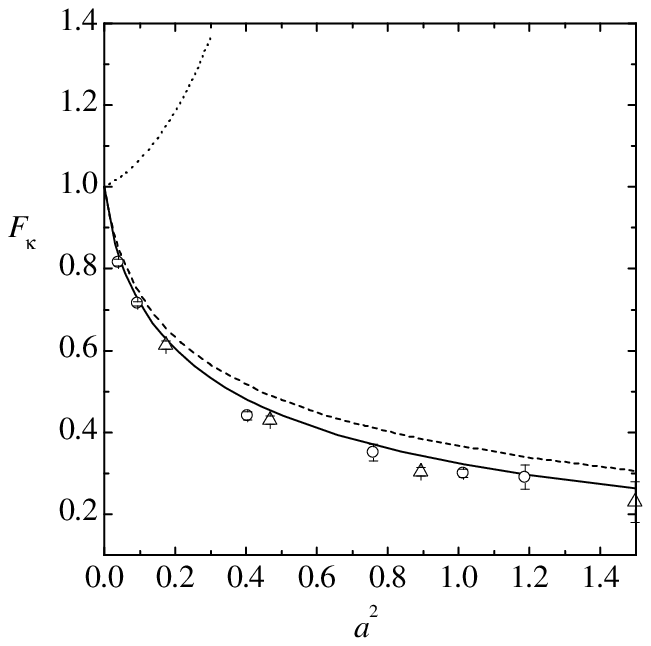}}
\caption{
Same as in Fig.\ \protect\ref{fig7}, but for the thermal conductivity
function $F_\kappa$.
 \label{fig9}}
\end{center}
\end{figure}

The most important transport coefficient is the nonlinear
viscosity represented by the function $F_\eta(a)$. According to
Fig.\ \ref{fig8}, the three theories retain the qualitative trends
of the simulation data, namely the decrease of $F_\eta$ with
increasing $a$ (shear thinning effect). In general, however, the
kinetic models (especially the BGK model) have a better agreement
than the Grad method. It is also interesting to remark that a
slight influence of the interaction potential seems to exist, the
shear thinning effect being a little bit more significant for hard
spheres than for Maxwell molecules. The nonlinear thermal
conductivity $F_\kappa(a)$ is plotted in Fig.\ \ref{fig9}. It is
quite apparent that  Grad's solution does not capture even the
qualitative shape of $F_\kappa$, as was already noted in the case
of hard disks.\cite{MG98} Again, the kinetic models present a good
agreement, especially in the case of the ES model.

\begin{figure}
\begin{center}
\parbox{0.5\textwidth}{
\epsfxsize=\hsize \epsfbox{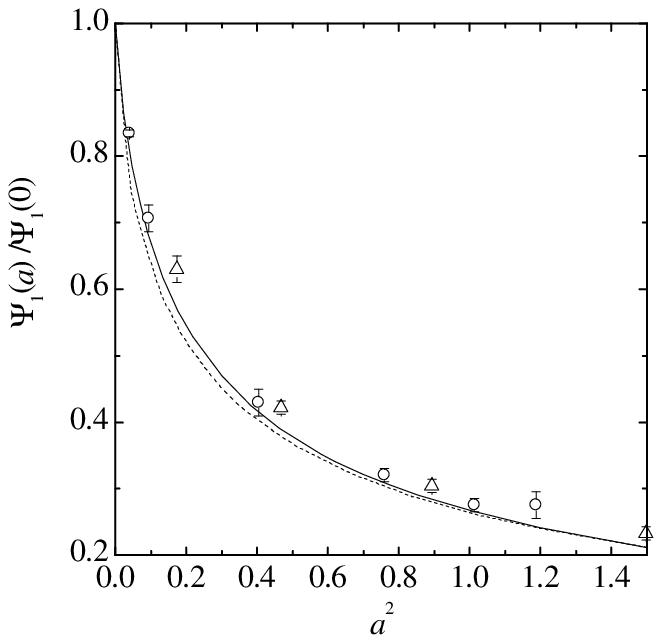}}
\caption{
Same as in Fig.\ \protect\ref{fig7}, but for the first viscometric
function $\Psi_1$, relative to the Burnett value $\Psi_1(0)$.
Note that the Grad prediction is not plotted.
 \label{fig10}}
\end{center}
\end{figure}

\begin{figure}
\begin{center}
\parbox{0.5\textwidth}{
\epsfxsize=\hsize \epsfbox{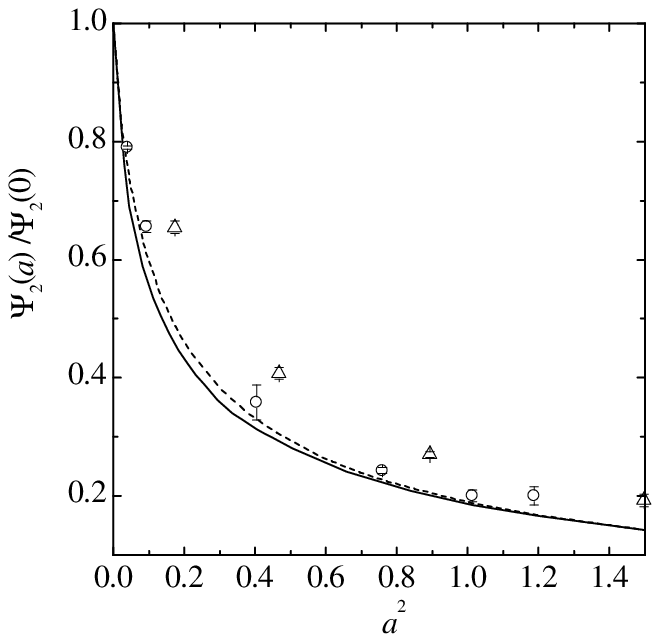}}
\caption{
Same as in Fig.\ \protect\ref{fig7}, but for the second viscometric
function $\Psi_2$, relative to the Burnett value $\Psi_2(0)$.
Note that the Grad prediction is not plotted.
 \label{fig11}}
\end{center}
\end{figure}

\begin{figure}
\begin{center}
\parbox{0.5\textwidth}{
\epsfxsize=\hsize \epsfbox{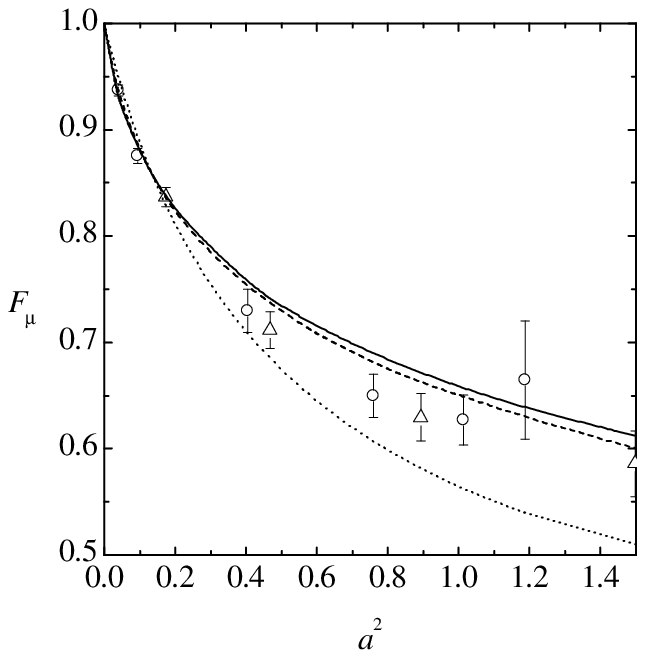}}
\caption{
Same as in Fig.\ \protect\ref{fig7}, but for the friction function $F_\mu$.
 \label{fig12}}
\end{center}
\end{figure}

\begin{figure}
\begin{center}
\parbox{0.5\textwidth}{
\epsfxsize=\hsize \epsfbox{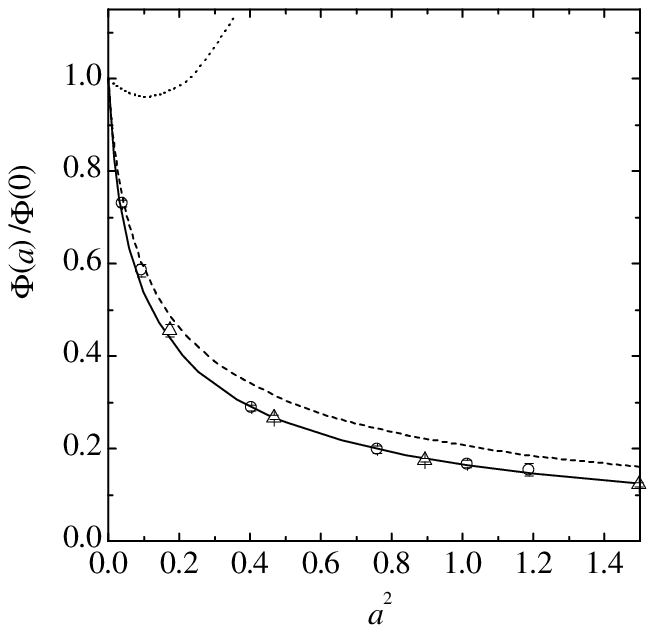}}
\caption{
Same as in Fig.\ \protect\ref{fig7}, but for the cross coefficient $\Phi$, relative to the Burnett value $\Phi(0)$.
 \label{fig13}}
\end{center}
\end{figure}

Normal stress differences are characterized by the viscometric functions
$\Psi_{1,2}(a)$. These quantities are well described by the kinetic models,
as shown in Figs.\ \ref{fig10} and \ref{fig11}. At a quantitative level, the
agreement is better in the case of Maxwell molecules. In fact, the
viscometric functions, especially the second one, exhibit a certain influence of the potential.
Although the functions $\Psi_{1,2}(a)$ were not evaluated from the Grad method in
Refs.\ \onlinecite{RC97} and \onlinecite{RC98}, the friction function $F_\mu(a)$ was calculated. This quantity is plotted in Fig.\ 
\ref{fig12}, showing an agreement between the theories and the simulation data similar to the one found in Fig.\ \ref{fig8} for the 
viscosity function.
The last transport coefficient is the cross-coefficient $\Phi$ defined by
Eq.\ (\ref{2.18}).
The comparison with simulation results for this quantity is a stringent
test of the theories, since it is a generalization of a Burnett
coefficient that measures the component of the heat flux orthogonal to the
thermal gradient.
Figure \ref{fig13} shows that, as happened with the thermal conductivity
function $F_\kappa(a)$, the Grad method gives a wrong prediction for the
shear-rate dependence of $\Phi(a)$.
On the other hand, the kinetic models describe fairly well the
nonlinear behavior of this function. In the case of the ES model, the agreement with the simulation data is practically perfect.

\subsection{Velocity distribution function}
\begin{figure}
\begin{center}
\parbox{0.5\textwidth}{
\epsfxsize=\hsize \epsfbox{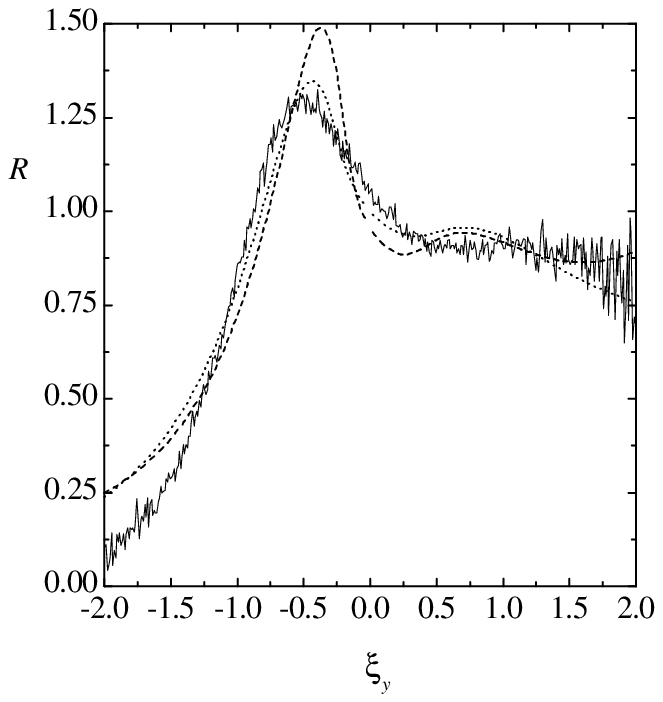}}
\caption{
Marginal velocity distribution function, relative to the local equilibrium  distribution, for Maxwell molecules at $y=0.5 L$ in the
case $a=0.636$ and $\epsilon=-0.272$. The solid line is
obtained from a DSMC simulation, while the dashed line
is the theoretical ES distribution and the dotted line is the theoretical BGK distribution.
 \label{fig14}}
\end{center}
\end{figure}

\begin{figure}
\begin{center}
\parbox{0.5\textwidth}{
\epsfxsize=\hsize \epsfbox{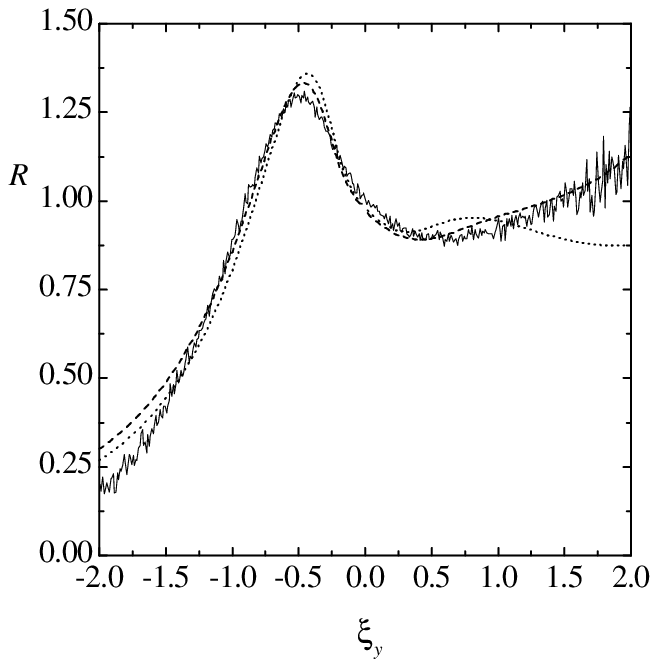}}
\caption{
Same as in Fig.\ \protect\ref{fig13}, but for hard spheres in the case $a=0.419$ and $\epsilon=-0.195$.
 \label{fig15}}
\end{center}
\end{figure}

Apart from  the transport coefficients, the kinetic models provide
the velocity distribution function. In the case of the BGK model,
the solution is given by Eq.\ (\ref{2.28}), while the ES
distribution function can be found in Refs.\
\onlinecite{GLH97} and \onlinecite{MG98}. In order to assess their reliability, we
have computed in the simulations the marginal distribution
(\ref{3.19.2}) at the layer $y/L=0.5$. The ratio $R(\xi_y)\equiv
\varphi(\xi_y)/[\pi^{-1/2}\exp(-\xi_y^2)]$, where $\varphi(\xi_y)$ is defined by Eq.\ (\ref{3.19.2}), is a measure of the
departure of the distribution function from the local equilibrium.
This quantity is plotted in Fig.\ \ref{fig14} for Maxwell
molecules in the case of  a reduced shear rate $a=0.636$ and a reduced
(local) thermal gradient $\epsilon=-0.272$. Both theories capture
the main features of the actual distribution. While the BGK
distribution exhibits a better agreement near the maximum (around
$\xi_y\simeq -0.5$), the ES distribution seems to describe better
the region $\xi_y\gtrsim 1$. The case of hard spheres is
illustrated in Fig.\ \ref{fig15}, which corresponds to $a=0.419$
and $\epsilon=-0.195$. In this case, the ES model shows a superiority over the
BGK model both near the maximum and for large positive velocities.

\section{Concluding remarks}
\label{sec5}
This paper has dealt with the steady planar Couette flow in a dilute gas
beyond the scope of the Navier-Stokes description.
This nonlinear problem had been previously studied by means of kinetic
theory tools, such as the Grad method, and the BGK and ES kinetic models.
These theories predict momentum and heat fluxes characterized by five
shear-rate dependent
generalized transport coefficients: a viscosity function $F_\eta(a)$, Eq.\
(\ref{2.14}), a thermal conductivity function $F_\kappa(a)$, Eq.\
(\ref{2.15}), two viscometric functions $\Psi_{1,2}(a)$, Eqs.\
(\ref{2.16}) and (\ref{2.17}), and a cross
coefficient $\Phi(a)$, Eq.\ (\ref{2.18}).
The main motivation of our study has been to perform DSMC simulations for
Maxwell molecules and hard spheres
in order to assess the reliability of the above theories in the
non-Newtonian regime.
Since we have been interested in the bulk properties, we have used
``nonequilibrium'' boundary conditions to inhibit the influence of
finite-size effects.

An important outcome of the simulation results is that, as predicted by the
kinetic theories considered here,
the shear-rate dependence of the transport coefficients is practically
insensitive to the interaction potential when the quantities are properly
nondimensionalized.
In particular, the actual shear rate has been reduced with respect to an
effective collision frequency defined from the Navier-Stokes shear viscosity coefficient.
The simulation results show, however, that the second viscometric function
presents a non negligible influence of the interaction model, so that the normal
stress difference $P_{zz}-P_{yy}$ is smaller for Maxwell molecules than for
hard spheres.
The comparison with the theoretical predictions shows that the kinetic
models give a fairly good description of the five transport coefficients.
On the other hand,  the Grad method yields a shear viscosity in qualitative
agreement with the simulations but
it dramatically fails for the coefficients measuring the heat flux. This is
basically due to the truncation scheme of the Grad method at the level of the
heat flux.
The physical idea behind a kinetic model is quite different, since it
consists of replacing the true Boltzmann collision operator by a simple
relaxation term but otherwise all the velocity moments are taken into
account.
As a consequence, while in the Grad method one has to solve a closed set of
coupled differential equations for the moments, in the case of the kinetic
model one gets the velocity distribution function and determines the fluxes
from it.

In the ES kinetic model the reference distribution function
appearing in the collision operator is an anisotropic Gaussian
parameterized by the pressure tensor. This allows the model to
give the correct Prandtl number $\text{Pr}=\frac{2}{3}$ but at the
expense of complicating its mathematical structure. In the case of
the BGK model, however, the reference distribution is that of
local equilibrium but the model leads to $\text{Pr}=1$. The
agreement with simulation of the ES model is generally better than
that of the BGK model, especially in the case of $F_\kappa$ and
$\Phi$. In spite of this, it is fair to say that the performance
of the BGK model is quite good, given its simplicity relative to
that of the ES model. Finally, the results reported in this paper
clearly shows the usefulness of kinetic models to analyze
nonlinear transport phenomena in the Couette flow problem. This
complements previous conclusions drawn from other nonlinear
problems, such as the uniform shear flow and the Fourier flow.

\acknowledgments 
The authors acknowledge partial support from the DGES (Spain)
through grant No.\ PB97-1501 and from the Junta de Extremadura
(Fondo Social Europeo) through grant No.\ IPR99C031.

\appendix
\section*{Theoretical expressions for the transport coefficients}
\label{appA}
In this Appendix we list the explicit shear-rate dependence of the
dimensionless transport coefficients defined in Sec.\ \ref{sec2}, according
to the Grad method, the BGK model, and the ES model.
\subsection{The Grad method}
From the Appendix  of Ref.\ \onlinecite{RC97}, corrected in Ref.\
\onlinecite{RC98}, one has
\beq
F_\eta(a)=\frac{2}{1+\frac{72}{25}a^2+\Delta(a)},
\label{A1}
\eeq
\beq
F_\kappa(a)=\frac{4}{1-\frac{216}{25}a^2+3\Delta(a)},
\label{A2}
\eeq
\beq
\Phi(a)=-7\frac{1-\frac{36}{125}a^2}{1+\frac{6}{5}a^2+(1-\frac{63}{25}a^2)
\Delta(a)},
\label{A3}
\eeq
where $\Delta(a)\equiv \sqrt{1+\frac{116}{25} a^2-\frac{864}{625}a^4}$.
The viscometric functions are not evaluated in Refs.\ \onlinecite{RC97} and \onlinecite{RC98},
although  the friction function is provided. It is given by
\beq
F_\mu(a)=\frac{2}{1+\frac{12}{25}a^2+\Delta(a)}.
\label{A4}
\eeq
Note that $F_\eta(a)$ and $F_\mu(a)$ become meaningless for $a^2\geq 25(\sqrt{1057}+29)/432\simeq 3.56$, while $F_\kappa(a)$ and 
$\Phi(a)$ are unphysical for $a^2\geq 50/63\simeq 0.79$.
For small shear rates, the above transport coefficients behave as $F_\eta\simeq 1-\frac{13}{5}a^2$, $F_\kappa\simeq 1+\frac{21}{50}a^2$, 
$\Phi\simeq -\frac{7}{2}(1-\frac{197}{250}a^2)$, and $F_\mu\simeq 1-\frac{7}{5}a^2$.

\subsection{The BGK model}
The derivation of the transport coefficients from the BGK model \cite{BSD87}
implies the resummation of asymptotic series by means of the Borel method.
As a consequence, the results are expressed in terms of the functions
$F_r(x)$ defined by the recurrence relation $F_r(x)=[(d/dx)x]^rF_0(x)$, where
\beq
F_0(x)=\frac{2}{x}\int_0^\infty dt\, t e^{-t^2/2}K_0(2t^{1/2}/x^{1/4}),
\label{A5}
\eeq
$K_0$ being the zeroth-order modified Bessel function.
The curvature parameter $\gamma(a)$ is given by the solution to the
following implicit equation
\beq
a^2=\gamma \frac{3F_1+2F_2}{F_1},
\label{A6}
\eeq
where the functions $F_r$ are evaluated at $x=\gamma$.
Next, the transport coefficients are expressed in terms of $a$  and
$F_r(\gamma)$ as
\beq
F_\eta(a)=F_0,
\label{A7}
\eeq
\beq
F_\kappa(a)=\frac{F_0}{5}\frac{3F_1+2F_2}{F_1},
\label{A7bis}
\eeq
\beq
\Psi_1(a)=-2F_1\frac{3F_1+4F_2}{3F_1+2F_2},
\label{A8}
\eeq
\beq
\Psi_2(a)=4\frac{F_1F_2}{3F_1+2F_2},
\label{A9}
\eeq
\beq
\Phi(a)=-\left[5F_2+2F_3+a^2(F_2+5F_3+8F_4+4F_5)\right].
\label{A10}
\eeq
For small shear rates, one gets $F_\eta\simeq 1-\frac{18}{5}a^2$, $F_\kappa\simeq 1-\frac{162}{25}a^2$, $\Psi_1\simeq 
-\frac{14}{5}(1-\frac{1476}{175}a^2)$, $\Psi_2\simeq \frac{4}{5}(1-\frac{288}{25}a^2)$, $\Phi\simeq -\frac{14}{5}$, and $F_\mu\simeq 
1-\frac{12}{5}a^2$.

\subsection{The ES model}
In Refs.\ \onlinecite{GLH97} and \onlinecite{MG98} the solution of the ES model is worked out
keeping the Prandtl number $\text{Pr}$ as a free parameter. Here we
particularize the results to the correct value $\text{Pr}=\frac{2}{3}$.
It is convenient to express the transport coefficients in terms of an
auxiliary parameter $\beta$, defined as the solution of the  implicit
equation
\beq
a^2=\frac{4\beta}{9}\frac{[2\beta(F_1+2F_2)-3]^2[3F_1+2F_2-2\beta
F_1(F_1+2F_2)]}{F_0^2[2\beta(F_1+F_2)-3]+F_1[2\beta(F_1+2F_2)-3]^2},
\label{A11}
\eeq
where now  the functions $F_r$ are evaluated at $x=\beta$. The relationship
between the curvature parameter $\gamma$ and $\beta$ is
\beq
\gamma(a)=\frac{2}{9}\beta[3-2\beta(F_1+2F_2)].
\label{A12}
\eeq
The transport coefficients are
\beq
F_\eta(a)=\frac{9F_0}{[2\beta(F_1+2F_2)-3]^2},
\label{A13}
\eeq
\beq
F_\kappa(a)=\frac{a^2}{5\gamma(a)}F_\eta(a),
\label{A14}
\eeq
\beq
\Psi_1(a)=-\frac{12\beta}{a^2}\frac{3F_1+4F_2-2\beta
F_1(F_1+2F_2)}{(3- 2\beta F_1)[3-2\beta F_1(F_1+2F_2]},
\label{A15}
\eeq
\beq
\Psi_2(a)=\frac{24\beta}{a^2}\frac{F_2}{(3- 2\beta F_1)[3-2\beta
F_1(F_1+2F_2]},
\label{A16}
\eeq
\beqa
\Phi(a)&=& \frac{1}{40} C_1
\left\{a^2[C_1^4F_0^3(F_1+2F_2)-9C_1^3F_0^2(F_2+2F_3)+54C_1^2F_0(F_2+4(F_3+F_4))
\right.\nonumber\\
&&-108C_1(F_2+5F_3+4(2F_4+F_5))]-C_3(C_1F_0F_1-6F_2)+4C_1^2F_0(F_1+2F_2)
\nonumber\\
&&\left.+4C_1[C_2F_0F_1-6(F_2+2F_3)]-24C_2F_2\right\}.
\label{A17}
\eeqa
In Eq.\ (\ref{A17}),
\beq
C_1\equiv \frac{3}{3-2\beta(F_1+2F_2)},
\label{A18}
\eeq
\beq
C_2\equiv \frac{3}{3-2\beta F_1},
\label{A19}
\eeq
\beq
C_3\equiv 3a^2C_1^3F_0^2+4C_2\{2\beta[C_1(F_1+4F_2)+3F_1]
-3C_1\}
.
\label{A20}
\eeq
For small shear rates, one has $F_\eta\simeq 1-\frac{21}{5}a^2$, $F_\kappa\simeq 1-\frac{197}{25}a^2$, $\Psi_1\simeq 
-\frac{14}{5}(1-\frac{2126}{175}a^2)$, $\Psi_2\simeq \frac{4}{5}(1-\frac{413}{25}a^2)$, $\Phi\simeq -\frac{7}{2}$, and $F_\mu\simeq 
1-3a^2$.

%\end{document}
\newpage

\begin{references}
\bibitem[a)]{chema}Electronic mail: jmm@unex.es
\bibitem[b)]{andres}Electronic mail: andres@unex.es
\bibitem[c)]{vicente}Electronic mail: vicenteg@unex.es

\bibitem{C90}
 C. Cercignani, {\em Mathematical Methods in Kinetic
Theory} (Plenum Press, New York, 1990).

\bibitem{CC70}
S. Chapman and T. G. Cowling, {\em The Mathematical Theory of
Nonuniform Gases} (Cambridge University Press, Cambridge, 1970).

\bibitem{MN80}
N. K. Makashev and V. I. Nosik, ``Steady-state Couette flow (with heat transfer) of a gas of Maxwellian molecules,'' Dokl. Akad. Nauk 
SSSR {\bf
253}, 1077 (1980) [Sov. Phys. Dokl. {\bf 25}, 589 (1981)]; V. I. Nosik, ``Degeneration of the Chapman-Enskog expansion in one-dimensional 
motions of Maxwellian molecule gases,'' in
{\em Rarefied Gas Dynamics 13},
edited by O. M. Belotserkovskii, M. N. Kogan, S. S. Kutateladze, and A. K.
Rebrov (Plenum Press, New York, 1983), Vol.\
1, pp.\ 237--244.

\bibitem{TS95}
M. Tij and A. Santos, ``Combined heat and momentum transport in a dilute gas,'' Phys. Fluids {\bf 7}, 2858 (1995).

\bibitem{BSD87}
 J. J. Brey, A. Santos, and J. W. Dufty, ``Heat and momentum transport far from equilibrium,''
Phys. Rev. A {\bf 36}, 2842 (1987).

\bibitem{KDSB89}
 C. S. Kim, J. W. Dufty, A. Santos, and J. J. Brey, ``Analysis of nonlinear transport in Couette flow,'' Phys. Rev. A {\bf 40},
7165 (1989).

\bibitem{GLH94}
V. Garz\'o and M. L\'opez de Haro, ``Kinetic model for heat and momentum transport,'' Phys. Fluids {\bf 6},
3787 (1994).

\bibitem{GLH97}
V. Garz\'o and M. L\'opez de Haro, ``Nonlinear transport for a dilute gas in  steady Couette flow,'' Phys. Fluids {\bf 9},
776 (1997).

\bibitem{MG98}
J. M. Montanero and V. Garz\'o, ``Nonlinear Couette flow in a dilute gas: Comparison between theory and molecular-dynamics simulation,'' 
Phys. Rev. E {\bf 58}, 1836 (1998).

\bibitem{RC97}
D. Risso and P. Cordero, ``Dilute gas Couette flow: Theory and molecular-dynamics simulation,'' Phys. Rev. E {\bf 56}, 489 (1997).

\bibitem{Bird94}
 G. A. Bird, {\em Molecular Gas Dynamics and the Direct
Simulation of Gas Flows} (Clarendon, Oxford, 1994).

\bibitem{TTKB98}
 R. Tehver, F. Toigo, J. Koplik, and J. R. Banavar, ``Thermal walls in computer simulations,'' Phys. Rev. E
{\bf 57}, R17 (1998).

\bibitem{MASG94}
J. M. Montanero, M. Alaoui, A. Santos, and V. Garz\'o, ``Monte Carlo simulation of the Boltzmann equation for steady Fourier flow,'' 
Phys.
Rev. E {\bf 49}, 367 (1994).

\bibitem{SG95}
A. Santos and V. Garz\'o, ``Exact non-linear transport from the Boltzmann equation,'' in {\em Rarefied Gas Dynamics 19},
edited by J. Harvey and G. Lord (Oxford University Press, Oxford, 1995), Vol.\ 1, pp.\ 13--22.

\bibitem{McL89}
J. A. McLennan, {\em Introduction to Nonequilibrium Statistical
Mechanics} (Prentice-Hall, Englewood Cliffs, NJ, 1989).

\bibitem{RC98}
D. Risso and P. Cordero, ``Erratum: Dilute gas Couette flow: Theory and molecular-dynamics simulation [Phys. Rev. E {\bf 56}, 489 
(1997)],'' Phys. Rev. E {\bf 57}, 7365 (1998).

\bibitem{DB77}
J. R. Dorfman and H. van Beijeren, ``The kinetic theory of gases,'' in {\em Statistical Mechanics}, Part B, edited by B. J. Berne 
(Plenum, New York), pp.\ 65--179.

\bibitem{W93}
D. C. Wadsworth, ``Slip effects in a confined rarefied gas: I: Temperature slip,'' Phys. Fluids A {\bf 5}, 1831 (1993).

\bibitem{instab}
J. M. Montanero, A. Santos, M. Lee, J. W. Dufty, and J. F. Lutsko, ``Stability of uniform shear flow,'' Phys. Rev. E {\bf 57}, 546 
(1998).

\bibitem{MS96}
In the case of very small Knudsen numbers, it is known that the DSMC method applied to the shear flow problem reproduces the NS shear 
viscosity $\eta_0$. See, for instance, J. M. Montanero and A. Santos, ``Monte Carlo simulation method for the Enskog equation,'' Phys. 
Rev. E {\bf 54}, 438 (1996); ``Simulation of the Enskog equation {\em \`a la\/} Bird,'' Phys. Fluids {\bf 9}, 2057 (1997).



\end{references}
\end{document}